 \documentclass[authoryear,preprint,review,12pt]{elsarticle}

\usepackage{graphics}

\usepackage{graphicx}

\usepackage{epsfig}

\usepackage{amssymb}

\usepackage{amsthm}

\usepackage{lineno}

\journal{New Astronomy}

\begin{document}

\begin{frontmatter}



\title{Is KR\,Cygni a Triple Star System?}


\author{Esin Sipahi\corref{cor1}}
\ead{esin.sipahi@mail.ege.edu.tr}
\cortext[cor1]{Corresponding author}

\address{Ege University, Science Faculty, Department of Astronomy and Space Sciences, 35100 Bornova, \.{I}zmir, Turkey}

\begin{abstract}
New multi-color UBVR light curves of the eclipsing binary KR\,Cyg were obtained in 2005. Photometric solutions were derived using the Wilson-Devinney method. The result shows that KR\,Cyg is a near-contact binary system with a large effective temperature difference between the components, approximately 5230 K. All the times of minimum light were collected and combined with our observations obtained in 2010 and 2011. Analysing all the times of mid-eclipse, the period of quasi-sinusoidal variation is about $\sim$76 years. The periodic oscillation could be explained by the light-time effect due to a presumed third component.
\end{abstract}


\end{frontmatter}



\section{Introduction}

Near-contact binaries are very important source in understanding the formation and evolution of the binary systems. The members of these systems have orbital periods less than 1 day. Most of the systems contain a component at or near its Roche lobe. Theoretical studies suggest that the near-contact binaries could be in the intermediate-stage between a detached or semi-detached configuration (e.g. \citealt{Hil88, Sha90}). Therefore, these systems are important observational targets which could play a key role in understanding the evolutionary stages of interacting binaries.

KR\,Cyg (HD\,333645, $V=9^{m}.19$) was recognized as a variable star by \citet{Sch31a, Sch31b, Sch31c}. Its properties are relatively poorly known compared to those of other short-period binaries. The observations of \citet{Vet65} revealed the first photoelectric light curves and a period of $0^{d}.8451538$ for the system. \citet{Sha90} included KR\,Cyg in their list of near-contact eclipsing binaries. Its more detailed history until 2004 has been described by \citet{Sip04}. Unfortunately, no spectroscopic study exists. 

We followed the observations of the system and obtained new light curves in 2005. We report the new photometric observations, and the solutions of the light curves. In this paper, we investigate possible implications of the orbital period change of the system based on our new minima times, and on historical data collected from the literature. Both the period analysis and the light-curve analysis show that KR\,Cyg is probably a triple system. The system is important not only because it joins the group of near-contact binary systems \citep{Sha90}, but also because we show that it is part of a triple system with $\sim$76 $yr$ orbit and a mass function of $\sim0.00103 M_{\odot}$.

\section{Observations}

The observations were acquired with a High-Speed Three Channel Photometer attached to the 48 cm Cassegrain telescope at Ege University Observatory. The observations were continued in UBVR bands. Some basic parameters of the program stars are listed in Table 1. The names of the stars are listed in the first and second columns, while J2000 coordinates are listed in the third and fourth columns. The V magnitudes are in the next column, and the spectral types of the stars are listed in the last column. Although the variable and comparison stars are very close on the sky, differential atmospheric extinction corrections were applied. The atmospheric extinction coefficients were obtained from observations of the comparison stars on each night. Moreover, the comparison stars were observed with the standard stars in their vicinity and the reduced differential magnitudes, in the sense variable minus comparison, were transformed to the standard system using procedures out-lined by \citet{Har62}. The standard stars are used from the catalogues of \citet{Lan83, Lan92}. Furthermore, the dereddened colours of the system were computed. Heliocentric corrections were also applied to the times of the observations. To compute the standard deviations of observations, we used the standard deviations of the reduced differential magnitudes in the sense comparisons minus check stars for each night. In Figure 1, the light and colour curves of KR\,Cyg are shown. 

\begin{table*}
\centering
\caption{The data for variable, comparison, and check stars.}
\vspace{0.3cm}
\begin{tabular}{lccccc}
\hline\hline
Star	&	ID	&	Alpha (J2000)	&	Delta (J2000)	&	$m_{V}$ 	&	Sp.T.	\\
	&		&	($^{h}$ $^{m}$ $^{s}$)	&	($^{\circ}$ $^{\prime}$ $^{\prime\prime}$)	&	(mag)	&		\\
\hline
KR\,Cyg	&	HD\,333645	&	20 09 05.6	&	$+$30 33 01.3	&	9.19	&	A1V	\\
Comp. Star	&	HD\,191398	&	20 08 39.4	&	$+$30 20 15.2	&	9.01	&	A0V	\\
Check Star	&	HD\,333664	&	20 09 18.1	&	$+$30 13 39.4	&	9.58	&	A0	\\
\hline 
\end{tabular} 
\end{table*}

\begin{figure*}
\hspace{3.4 cm}
\includegraphics[width=15cm]{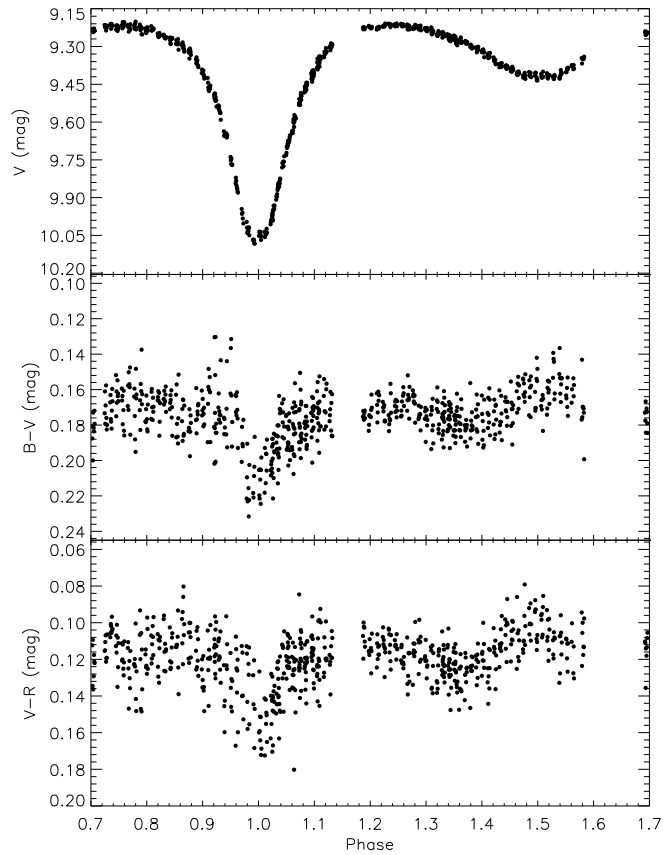}
\vspace{0.1 cm}
\caption{The V band light curve (upper panel) with both B-V (middle panel) and V-R (bottom panel) colour curves of KR\,Cyg.}
\label{Fig.1}
\end{figure*}

\section{Orbital Period Analysis}

In order to study of the orbital period of KR\,Cyg, all the available times of light minimum were collected from the literature. We also observed the system for two nights in 2010 and one night in 2011. Using these observations, three times of primary minimum were obtained. Finally, 156 visual, 19 photographic and 129 photoelectric times of mid-eclipse of KR\,Cyg were used to period study. New timings are listed in Table 2. The other minima times can be taken from the modern database of $O-C$ Gateway \citep{Pas06}. The $O-C$ residuals of KR\,Cyg were calculated using the following linear ephemeris;

\begin{center}
\begin{equation}
JD~(Hel.)~=~2455036.5239(8)~+~0^{d}.8451526(1)~\times~E.
\end{equation}
\end{center}

Then, the $(O-C)_{I}$ (upper panel) and $(O-C)_{II}$ (bottom panel) residuals were plotted against the epoch number ($E$) in Figure 2 where the times of photoelectric minima are marked by filled circle. The $(O-C)_{II}$ residuals shows that the orbital period of KR\,Cyg is changing in a long-term sinusoidal form, although the visual and photographic data display large scatter throughout the diagram. Assuming that the most likely cause of the cyclic variation in the $(O-C)_{II}$ residuals could be light-travel-time effect (LTTE) due to an unseen third body in the system. The following LTTE equation, as given by \citet{Irw59}, was fitted to the $(O-C)_{II}$ residuals of the eclipse timings of the system:

\begin{center}
\begin{equation}
\Delta t~=~\frac{a_{12}sini'}{c} \lbrace\frac{1-e'^{2}}{1+e'cos\nu'}sin(\nu'+\omega')+e'cos\omega'\rbrace
\end{equation}
\end{center}
where $\Delta t$ is the time advance/delay due to LTTE, $c$ is the speed of light, and $a_{12}$, $i'$, $e'$ and $\omega'$ are the semi-major axis, inclination, eccentricity and longitude of the periastron of the absolute orbit of the centre of mass of the eclipsing binary around the center-of-mass of the triple system, respectively. $\nu'$ is the true anomaly of the position of the eclipsing binary's mass-centre on this orbit. 

A weighted least-squares solution for two parameters ($T_{0}$ and $P$) of the linear ephemeris of KR\,Cyg and five parameters ($a_{12} sin i'$, $e'$, $\omega'$, $T'$ and $P'$) of the LTTE are presented in Table 3. The observational points and theoretical best fit are plotted against the epoch number in the bottom panel of Figure 2. Assuming that the orbit of the presumed third body is circular, we obtained the mass function as $f(M_{3}) = 0.00103M_{\odot}$ for the third body, using the Equation:

\begin{center}
\begin{equation}
f(m)~=~\frac{4\pi^{2}}{GT^{2}}\times(a_{12}sini')^{3}=\frac{(M_{3}~sini')^{3}}{(M_{1}+ M_{2}+ M_{3})^{2}}
\end{equation}
\end{center}
where $M_{1}$, $M_{2}$ and $M_{3}$ are the masses of the binary's components, and the third body, respectively. The mass of the third component can be estimated from Equation (3), depending on the orbital inclination. For example, the mass $M_{3,min}$ was estimated to be 0.33 $M_{\odot}$ for $i'^=90^\circ$. Here, the total mass of the eclipsing system was taken as 5.51 $M_{\odot}$. We calculated the third body masses at different inclination, $i'$ values, which are shown in Table 3. The parameters given in Table 3 suggest that KR\,Cyg has an eccentric orbit around the centre-of-mass of the third-body system with a period of  $\sim$76 years.

\begin{table*}
\centering
\caption{The new times of minimum light for KR\,Cyg.}
\vspace{0.3cm}
\begin{tabular}{ccccc}
\hline\hline
No.	&	HJD	&	Filter	&	Error	&	Min.	\\
\hline
1	&	2455386.4355	&	B	&	0.0001	&	I	\\
	&	2455386.4356	&	V	&	0.0001	&	I	\\
	&	2455386.4356	&	R	&	0.0001	&	I	\\
2	&	2455419.3832	&	B	&	0.0001	&	I	\\
	&	2455419.3831	&	V	&	0.0001	&	I	\\
3	&	2455691.5219	&	B	&	0.0001	&	I	\\
	&	2455691.5220	&	V	&	0.0001	&	I	\\
	&	2455691.5220	&	R	&	0.0001	&	I	\\
\hline 
\end{tabular} 
\end{table*}

\begin{figure*}
\hspace{2.5 cm}
\includegraphics[width=14cm]{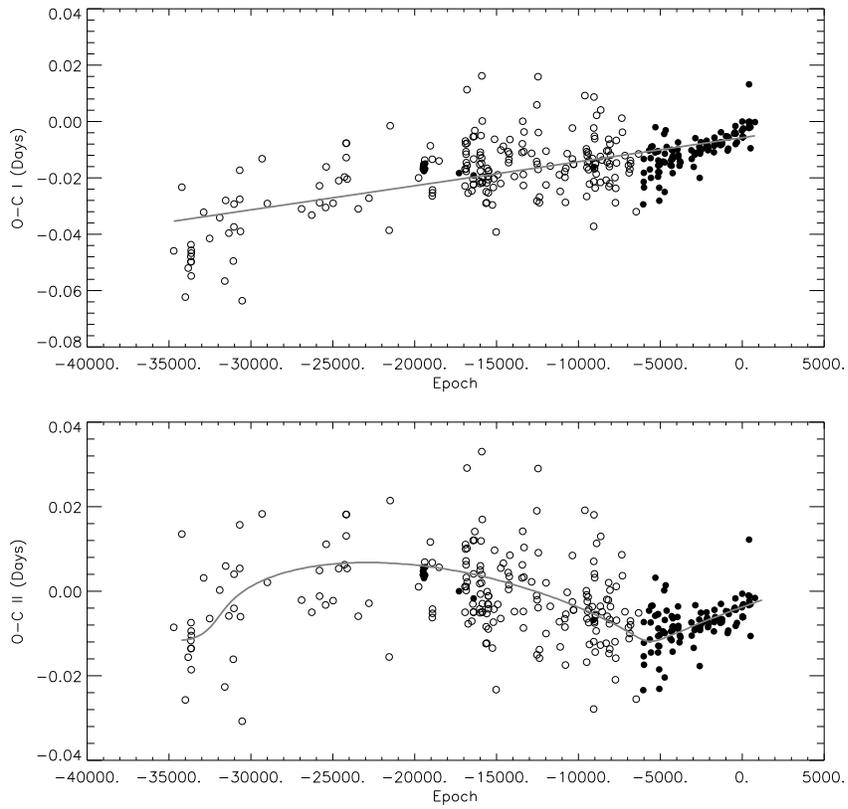}
\vspace{0.1 cm}
\caption{The $O-C$ variation of KR\,Cyg (The photoelectric minima times were marked by the filled circles).}
\label{Fig.2}
\end{figure*}

\begin{table*}
\centering
\caption{The parameters derived from $O-C$ analysis of KR\,Cyg.}
\vspace{0.3cm}
\begin{tabular}{lr}
\hline\hline
Parameter	&  Value	\\
\hline
$T_{0}$ (HJD)	&	2455036.5239$\pm$0.0008	\\
$P$ (day)	&	0.8451526$\pm$0.0000001	\\
$K$ (day)	&	0.010$\pm$0.001	\\
$P'$ (year) 	&	 75.6$\pm$0.6	\\
$e'$ 	&	 0.51$\pm$0.03	\\
$\omega'$ (deg) 	&	28.65$\pm$0.36	\\
$a_{12} sin i'$ (AU) 	&	1.8044$\pm$0.0041	\\
f(m) ($M_{\odot}$)	&	0.00103$\pm$0.00009	\\
$M_{3}$($i'$=$15^\circ$) & 1.42 $M_{\odot}$ \\
$M_{3}$($i'$=$30^\circ$) & 0.68 $M_{\odot}$ \\
$M_{3}$($i'$=$45^\circ$) & 0.48 $M_{\odot}$ \\
$M_{3}$($i'$=$60^\circ$) & 0.38 $M_{\odot}$ \\
$M_{3}$($i'$=$75^\circ$) & 0.34 $M_{\odot}$ \\
$M_{3}$($i'$=$90^\circ$) & 0.33 $M_{\odot}$ \\
\hline 
\end{tabular} 
\end{table*}

\section{Light Curve Analysis} 

Photometric analysis of KR\,Cyg was carried out using the PHOEBE V.0.31a software \citep{Prs05}. The method used in the PHOEBE V.0.31a software depends on the method used in the version 2003 of the Wilson-Devinney Code \citep{Wil71, Wil90}. The UBVR light curves were analysed simultaneously assuming the "detached" and "semi-detached binary, secondary star fills Roche lobe" configurations. In the process of the computation, we initially adopted the following fixed parameters: the mean temperature of the primary component ($T_{1}$), the linear limb-darkening coefficients of $x_{1}$ and $x_{2}$ for various bands, the gravity-darkening exponents of $g_{1}$, $g_{2}$ \citep{Luc67} and the bolometric albedo coefficients of $A_{1}$, $A_{2}$ \citep{Ruc69}. The commonly adjustable parameters employed are the orbital inclination ($i$), the mean temperature of the secondary component ($T_{2}$), the potentials of the components ($\Omega_{1}$ and $\Omega_{2}$) and the monochromatic luminosity of the primary component ($L_{1}$). The third light ($L_{3}$) was used also as free parameter to check for the third light contribution suggested by the period analysis.

We determined the colours of the system at phase 0.75 as follow: $U-B=0^{m}.101$, $B-V=0^{m}.156$, $V-R=0^{m}.128$. We derived dereddened colours as $(U-B)_{0}=-0^{m}.400$, $(B-V)_{0}=-0^{m}.12$. Then, we took JHK magnitudes of the system ($J=9^{m}.278$, $H=9^{m}.169$, $K=9^{m}.098$) from the 2MASS Catalogue \citep{Cut03}. Using these magnitudes, we derived dereddened colours as a $(J-H)_{0}=0^{m}.109$ and $(H-K)_{0}=0^{m}.071$ for the system. Using the calibrations given by Tokunaga (2000), we derived the temperature of the primary component as 11400 K and 11800 K depending on the UBV and JHK dereddened colours, respectively. Both of them indicate the same spectral type which is B8V. We adopted the mean temperature of the primary component as 11400 K for the light curve analyse. To find a reliable photometric mass ratio, the solutions are obtained for a series of fixed values of the mass ratio from $q=0.30$ to 0.6 in increments of 0.05. The sum of the squared residuals, ($\Sigma res^{2}$), for the corresponding mass ratios are plotted in Figure 3, where the lowest value of ($\Sigma res^{2}$) was found at about $q=0.45$. The photometric elements for the mass-ratio of 0.45 are listed in Table 4 and corresponding light curves are plotted in Figure 4 as continuous line. Moreover, the 3D model at phase 0.2 and Roche geometry of the system is displayed in Figure 5. Using the contribution of the third body to the total light of the system in U, B, V and R band, we calculated the instrumental $U-B$, $B-V$ and $V-R$ color indices of the third component as 0.027, $-0.022$ and $-0.163$, respectively. These color indices that the third body could be accepted as a normal star. Using the Mass-Luminosity relation of the main sequence stars, $L \alpha M^{3.5}$, we calculated the values of the third body luminosities. We compared these luminosities with the contribution of the third component in the light curve analysis and we obtained the mass of the third body as 0.48 $M_{\odot}$, corresponding to the inclination value $i'=45^\circ$. 

Although there is no available radial velocity curve of the system, we try to estimate the absolute parameters of the components. Considering its spectral type, we estimate the mass of the primary component from \citet{Tok00} as approximately 3.8 $M_{\odot}$, and the mass of the secondary component is calculated from the estimated mass ratio of the system. Using Kepler's third law, we calculate the semi-major axis ($a$), and then the mean absolute radii of the components. All the estimated absolute parameters are listed in Table 5. In Figure 6, we plot the components in the mass-radius plane. The Mass-Radius relationships for the ZAMS and TAMS model \citep{Pol98} also shown by the line and dotted line, respectively. In the figure, the open circle represents the secondary component, while the filled circle represents the primary component. Both of the components are located in the main-sequence band.

\begin{figure*}
\hspace{3 cm}
\includegraphics[width=12cm]{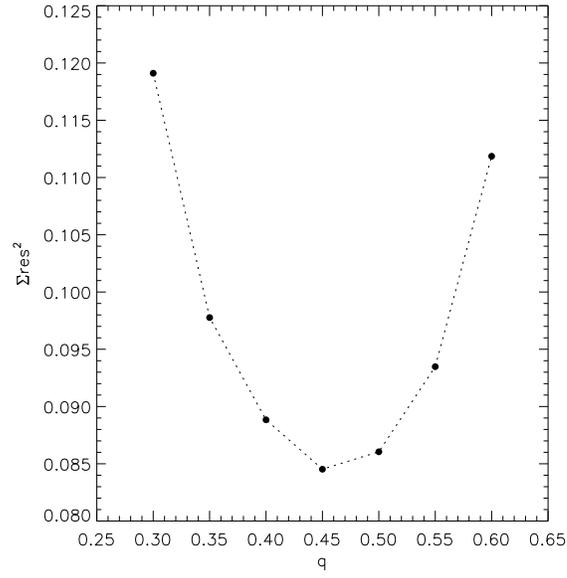}
\vspace{0.1 cm}
\caption{The variation of the sum of weighted squared residuals versus mass ratio.}
\label{Fig.3}
\end{figure*}

\begin{table*}
\centering
\caption{The parameters obtained from the light curve analysis.}
\vspace{0.3cm}
\begin{tabular}{cccc}
\hline\hline
	&	Detached	&&	Semi-detached	\\
\hline
Parameter	&	& Value &	\\
\hline
$q$	&	&0.45&	\\
$i$ ($^\circ$)	&	87.8$\pm$0.1	&&	87.74$\pm$0.04	\\
$T_{1}$ (K)	&	&11400&	\\
$T_{2}$ (K) 	&	6171$\pm$19	&&	6055$\pm$21	\\
$\Omega_{1}$	&	2.856$\pm$0.003	&&	2.872$\pm$0.002	\\
$\Omega_{2}$	&	2.718$\pm$0.003	&&	2.778$\pm$0.002	\\
L$_{1}$/L$_{T}$$(U)$	&	0.961$\pm$0.004	&&	0.969$\pm$0.003	\\
L$_{1}$/L$_{T}$$(B)$	&	0.951$\pm$0.004	&&	0.961$\pm$0.003	\\
L$_{1}$/L$_{T}$$(V)$	&	0.921$\pm$0.004	&&	0.933$\pm$0.003	\\
L$_{1}$/L$_{T}$$(R)$	&	0.891$\pm$0.004	&&	0.905$\pm$0.003	\\
L$_{3}$/L$_{T}$$(U)	$&	0.008$\pm$0.003	&&	0.005$\pm$0.002	\\
L$_{3}$/L$_{T}$$(B)	$&	0.009$\pm$0.002	&&	0.005$\pm$0.002	\\
L$_{3}$/L$_{T}$$(V)	$&	0.008$\pm$0.002	&&	0.005$\pm$0.002	\\
L$_{3}$/L$_{T}$$(R)	$&	0.007$\pm$0.002	&&	0.004$\pm$0.002	\\
$g_{1}$, $g_{2}$	&	&1.0, 0.5&	\\
$A_{1}$, $A_{2}$	&	&1.0, 0.5&	\\
$x_{1,bol}$, $x_{2,bol}$	&	&0.719, 0.643&	\\
$x_{1,U}$, $x_{2,U}$	&	&0.587, 0.869&	\\
$x_{1,B}$, $x_{2,B}$	&	&0.654, 0.828&	\\
$x_{1,V}$, $x_{2,V}$	&	&0.560, 0.745&	\\
$x_{1,R}$, $x_{2,R}$	&	&0.460, 0.653&	\\
$<r_{1}>$	&	0.4501$\pm$0.0006	&&	0.4456$\pm$0.0005	\\
$<r_{2}>$	&	0.3265$\pm$0.0008	&&	0.4187	\\
\hline
\end{tabular} 
\end{table*}

\begin{figure*}
\hspace{3.6 cm}
\includegraphics[width=18cm]{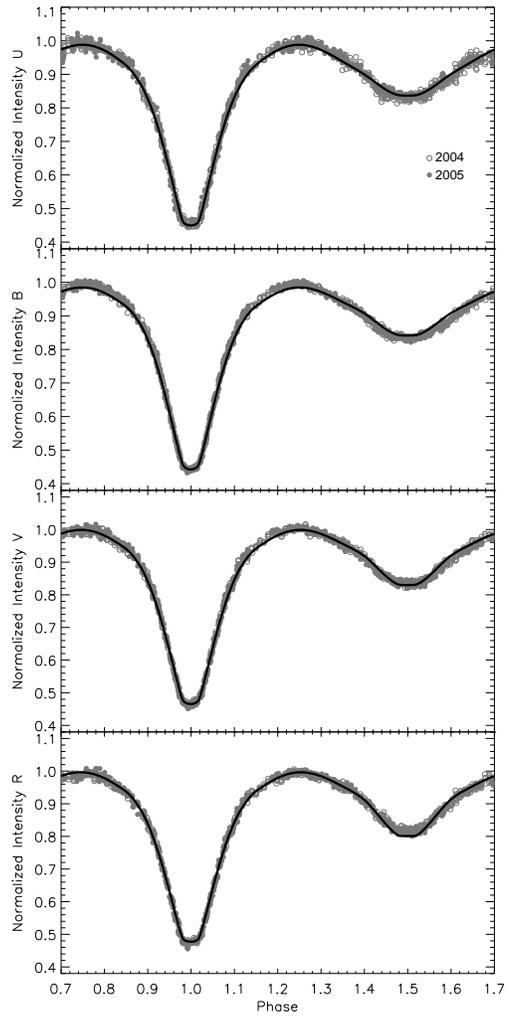}
\vspace{0.1 cm}
\caption{KR\,Cyg's light curves observed in UBVR bands and the synthetic curves derived from the best light curve solutions in each band.}
\label{Fig.4}
\end{figure*}

\begin{figure*}
\hspace{1.75 cm}
\includegraphics[width=10cm]{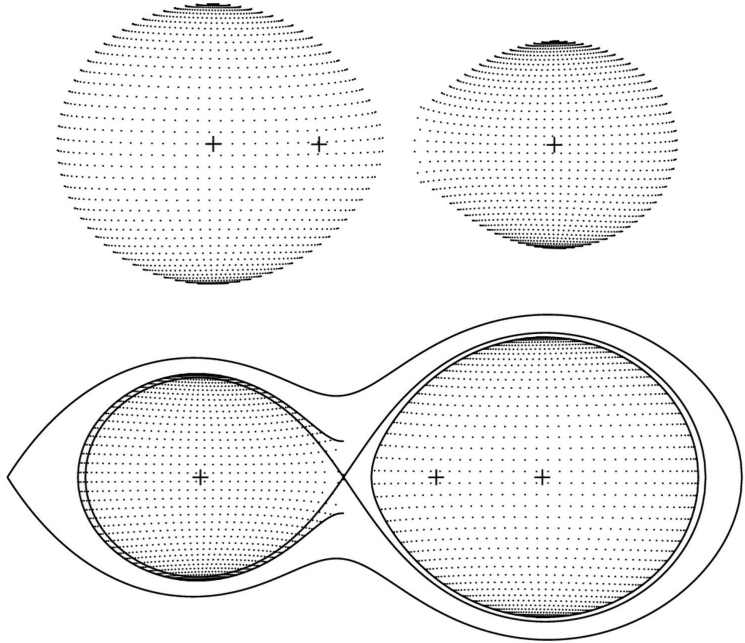}
\vspace{0.1 cm}
\caption{The 3D model at phase 0.2 (upper) and Roche geometry (bottom) of KR\,Cyg.}
\label{Fig.5}
\end{figure*}

\begin{figure*}
\hspace{3.5 cm}
\includegraphics[width=14cm]{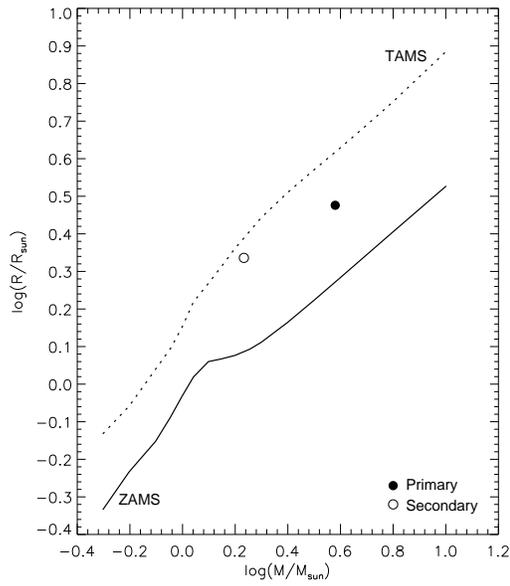}
\vspace{0.1 cm}
\caption{The places of the components of KR\,Cyg in the Mass-Radius plane. The continuous and dotted lines represent the ZAMS and TAMS theoretical model developed by \citet{Pol98}.}
\label{Fig.6}
\end{figure*}

\begin{table*}
\centering
\caption{The estimated absolute paramaters derived for KR\,Cyg.}
\vspace{0.3cm}
\begin{tabular}{lccc}
\hline\hline
Parameter 	&	Primary	&&	Secondary	\\
\hline
Mass ($M_{\odot}$) 	&	3.80	&&	1.71	\\
Radius ($R_{\odot}$) 	&	2.99	&&	2.17	\\
Luminosity ($L_{\odot}$) 	&	135.4	&&	6.10	\\
$M_{bol}$ (mag) 	&	-0.59	&&	2.78	\\
$log~g$	&	4.10	&&	4.00	\\
$d$ (pc)	&	&  452.16  &	\\
\hline 
\end{tabular} 
\end{table*}

\section{Summary}

We obtained new UBVR observations for KR\,Cyg and analysed the UBVR light curves to reveal a orbital parameters solution including its Roche configuration. A detailed analysis of the orbital period changes of KR\,Cyg was made using its $O-C$ diagram constructed from all available times of eclipse minima. The findings from this study are summarized as follows:

(i) New physical and geometrical parameters of components of the system have been derived. Orbital parameters indicate that KR\,Cyg is a near-contact binary system with a large temperature difference of approximately 5230 K.

(ii) KR\,Cyg is a semi-detached binary system, in which the primary (massive and hotter) component is slightly smaller than its Roche lobe, while the cool component is slightly greater than it. It exhibits a $\beta$ Lyrae type light curve, and fulfils most of the properties of the near-contact systems \citep{Sha90}. 

(iii) The system can be classified as near-contact binaries of FO Vir type. However, it must be noted that the existing observational material does not allow a precise classification for the system.

(iv) $O-C$ analysis indicates that KR\,Cyg is probably an interesting triple eclipsing system showing the slow LTTE caused by a third body orbiting with a period of $\sim$76 years. Furthermore, the significant contribution of the third-body found in the light curve analysis. Thus, it confirms that third body may produce the sinusoidal variation in the $O-C$ residuals.

(v) In future work, spectroscopic observations should be made to obtain radial velocity curves, which will allow a better discussion of the absolute dimensions and the evolutionary status of KR\,Cyg. New timings of this eclipsing binary are also necessary to improve the LTTE parameters derived in this paper.

\section*{Acknowledgments} The author acknowledge generous allotments of observing time at the Ege University Observatory.

\end{document}